\documentclass[10pt,a4paper]{article}

\usepackage[american]{babel}
\usepackage[utf8]{inputenc}
\usepackage[T1]{fontenc}

\usepackage{xspace}

\usepackage{graphicx}
\usepackage{hyperref}

\usepackage{url}

\usepackage{mathtools,amsmath,amsfonts,amsthm,paralist}

\begin{document}

\title{A Smoothed Particle Hydrodynamics Mini-App\\for Exascale}
\author{Aurélien Cavelan\\
\textit{University of Basel, Switzerland}\\
\textit{aurelien.cavelan@unibas.ch}
\and
Rubén M. Cabezón\\
\textit{University of Basel}\\
\textit{ruben.cabezon@unibas.ch}
\and
Michal Grabarczyk\\
\textit{University of Basel}\\
\textit{michalprzemyslaw.grabarczyk@unibas.ch}
\and
Florina M. Ciorba\\
\textit{University of Basel, Switzerland}\\
\textit{florina.ciorba@unibas.ch}
}

\maketitle

\newcommand{\flo}[1]{magenta{#1}}
\newcommand{\ac}[1]{blue{#1}}
\newcommand{\rc}[1]{red{#1}}

\newcommand{\changa}{ChaNGa\xspace}
\newcommand{\sphflow}{SPH-flow\xspace}
\newcommand{\sphynx}{SPHYNX\xspace}
\newcommand{\sphexa}{SPH-EXA~\cite{SPHEXA}\xspace}
\newcommand{\sphexaminiapp}{SPH-EXA mini-app\xspace}
\newcommand{\systemW}{Piz Daint\xspace}

\begin{abstract}
The Smoothed Particles Hydrodynamics (SPH) is a particle-based, meshfree, Lagrangian method used to simulate multidimensional fluids with arbitrary geometries, most commonly employed in astrophysics, cosmology, and computational fluid-dynamics (CFD). It is expected that these computationally-demanding numerical simulations will significantly benefit from the up-and-coming Exascale computing infrastructures, that will perform $10^{18}$ FLOP/s.
In this work, we review the status of a novel \sphexaminiapp, which is the result of an interdisciplinary co-design project between the fields of astrophysics, fluid dynamics and computer science, whose goal is to enable SPH simulations to run on Exascale systems.
The \sphexaminiapp merges the main characteristics of three state-of-the-art parent SPH codes (namely \changa, \sphflow, \sphynx) with state-of-the-art (parallel) programming, optimization, and parallelization methods.
The proposed \sphexaminiapp is a C++14 lightweight and flexible header-only code with no external software dependencies. Parallelism is expressed via multiple programming models, which can be chosen at compilation time with or without accelerator support, for a hybrid process+thread+accelerator configuration. 
Strong- and weak-scaling experiments on a production supercomputer show that the \sphexaminiapp can be efficiently executed with up 267 million particles and up to 65 billion particles in total on 2,048 hybrid CPU-GPU nodes.
\end{abstract}

\maketitle

\newpage
\section{Introduction}

The Smoothed Particle Hydrodynamics is a commonly used method to simulate the mechanics of continuum media. This method is a pure Lagrangian technique, particle-based, meshfree, and adequate for simulating highly distorted geometries and very dynamical scenarios, while conserving momentum and energy by construction. As such, it has been used in many different fields, including computational fluid dynamics, solid mechanics, engineering, nuclear fusion, astrophysics, and cosmology. In this work we present a review of the status of an SPH mini-app designed for Exascale. 

In most cases, actual SPH code implementations initially target a specific simulation scenario (or a subset). This means that in many cases these hydrodynamics codes are implemented with the physics needed to solve a specific problem in mind, while the parallelization, scaling, and resilience take a relevant role only later. This importance shift (or rebalancing) becomes relevant when addressing much larger problems that require much more computational resources, or when the computing infrastructures change or evolve. 
The philosophy behind the design of our \sphexaminiapp reflects the opposite. Knowing that the SPH technique is so ubiquitous, we developed the mini-app targeting the emerging Exascale infrastructures. We took into account the state-of-the-art SPH methodology, learnt from current production SPH codes, and implemented it having in mind the design characteristics for performance that would be desirable in a code that could potentially reach sustained ExaFLOP/s, namely scalability, adaptability, and fault tolerance. To reach this goal, we employed state-of-the-art computer science methodologies, explored different parallelization paradigms and their combinations, and used solid software development practices. All this within a close multi-directional interdisciplinary collaboration between scientists from fluid dynamics, astrophysics, cosmology, and computer science.

Lighter than production codes, mini-apps are algorithm-oriented and allow easy modifications and experiments \cite{BVH12}. Indeed, a single function can be evaluated using different strategies leading to different performance results, even if the physical result is unchanged. These evaluation strategies may rely on vectorization, node level multi-threading, or cross-node parallelism. Their efficiency also depends on platform configuration: presence of accelerators, generation of CPU, interconnection network fabric and topology, and others. Therefore, a mini-app is perfectly suitable as a portable code sandbox to optimize a numerical method, such as the SPH method, for Exascale. 

The \sphexaminiapp is a C++14 lightweight and flexible header-only code with no external software dependencies, that works by default with double precision data types. Parallelism is expressed via multiple programming models, which can be chosen at compilation time with or without accelerator support, for a hybrid node-core-accelerator configuration: MPI+OpenMP+OpenACC|OpenMP Target Offloading|CUDA. The \sphexaminiapp can be compiled with the GCC, Clang, PGI, Intel, and Cray (as long as the given compiler supports the chosen programming model) C/C++ compilers. The code is open-source and is freely available on GitHub under the MIT license at: \url{https://github.com/xxx/yyy}.

Weak-scaling experiments on a production supercomputer have shown that the \sphexaminiapp can execute with up to 65 billion particles\footnote{The largest cosmological SPH simulation to date performed with the parent codes is of the order of 25 billion particles \cite{ROMULUS}. Although, comparison is not fair due to different levels of physics included, it gives an order of magnitude reference.} on 2,048 hybrid CPU-GPU nodes at 67\% efficiency. These results are, therefore, very promising especially given that the efficiency of the code decreased very little when scaling from 512 to 2,048 nodes (see Section~\ref{subsec:scaling}).
Similar efficiency results are also reported in the strong-scaling results included in Figure~\ref{fig.strong-scaling}.


This paper is organized as follows. Section 2 presents a short overview of other representative mini-apps from scientific computing and describes the differences compared to skeleton applications. Section 3 is concentrated on the co-design aspects of this work. Section 4 includes all the details related to the present implementation of the \sphexaminiapp, including the implemented SPH version, the details on the employed parallel models and domain decomposition. Section 5 presents the results obtained regarding validation and verification, and the scaling experiments conducted. Finally, Section 6 discusses the next steps for the project and Section 7 presents the conclusions.

\section{Related Work}
\label{sec:related-work}

Mini-apps or proxy-apps have received great attention in recent years, 
with several projects being developed. 
The Mantevo Suite~\cite{HDCW09} devised at Sandia National Laboratory for high performance computing~(HPC) is one of the first large \mbox{mini-app} sets. It includes \mbox{mini-apps} that represent the performance of \mbox{finite-element} codes, molecular dynamics, and contact detection, to name a few. 

Another example is CGPOP~\cite{CGPOP}, a \mbox{mini-app} from oceanography, that implements a conjugate gradient solver to represent the bottleneck of the full Parallel Ocean Program application. CGPOP is used for experimenting with new programming models and to ensure performance portability.

At Los Alamos National Laboratory, MCMini~\cite{MCMini} was developed as a \mbox{co-design} application for Exascale research.
MCMini implements Monte Carlo neutron transport in OpenCL and targets accelerators and coprocessor technologies.

The CESAR Proxy-apps~\cite{CESAR} represent a collection of mini-apps belonging to three main fields: thermal hydraulics for fluid codes, neutronics for neutronics codes, and coupling and data analytics for \mbox{data-intensive} tasks.

The European ESCAPE project~\cite{ESCAPE} defines and encapsulates the fundamental building blocks (`Weather \& Climate Dwarfs') that underlie weather and climate services. This serves as a prerequisite for any subsequent co-design, optimization, and adaptation efforts. One of the ESCAPE outcomes is Atlas~\cite{Atlas}, a library for numerical weather prediction and climate modeling, with the primary goal of exploiting the emerging hardware architectures forecasted to be available in the next few decades. Interoperability across the variegated solutions that the hardware landscape offers is a key factor for an efficient software and hardware co-design~\cite{Schulthess2015}, thus of great importance when targeting Exascale systems.

In the context of parallel programming models, research has been focusing on the efficient use of intra-node parallelism, able to properly exploit the underlying communication system through a fine grain task-based approach, ranging from libraries (Intel TBB~\cite{TBB}) to language extensions (Intel Cilk Plus~\cite{CilkPlus} or OpenMP), to experimental programming languages with focus on productivity  (Chapel~\cite{Chapel}). Kokkos~\cite{Kokkos} offers a programming model, in C++, to write portable applications for complex manycore
architectures, aiming for performance portability. HPX~\cite{HPX} is a task-based asynchronous programming model that offers a solution for homogeneous execution of remote and local operations.

Similar to these works, the creation of a mini-app directly from existing codes rather than the development of a code that mimics a class of algorithms has been recently discussed~\cite{Messer2015}. A scheme to follow was proposed therein that must be adapted according to the specific field the parent code originates in. To maximize the impact of a mini-app on the scientific community, it is important to keep the \emph{build and execution system} simple, to not discourage potential users. The building should be kept as simple as a Makefile and the preparation of an execution run to a handful of command line arguments: 
``\textit{if more than this level of complexity seems to be required, it is possible that the resulting MiniApp itself is too complex to be human-parseable, reducing its usefulness.}" \cite{Messer2015}. 

The present work introduces the interdisciplinary co-design of an \sphexaminiapp with three parent SPH codes originating in the astrophysics academic community and the industrial CFD community. This represents a category not discussed in~\cite{Messer2015}. 


\section{Co-Design}

Being optimization critical to achieve the scalability needed to exploit Exascale computers, the long-term goal of the \sphexa is to provide a parallel, optimized, state-of-the-art implementation of basic SPH operands with classical test cases used by the SPH community. This can be implemented at different levels: employing state-of-the art programming languages, dynamic load balancing algorithms, fault-tolerance techniques, and optimized tools and libraries.

Interdisciplinary co-design and co-development~\cite{HDCW09} allow to adequately involve the developers of the parent codes (in our case \changa, \sphflow, and \sphynx), thus boosting and improving the design and the implementation of the \sphexaminiapp.
We employ an interdisciplinary co-design approach that goes beyond the classical hardware-software approach and \emph{holistically co-design} between the SPH \emph{application}, the \emph{algorithms} it employs (SPH method, distributed tree, finding neighbors, load balancing, silent data corruption detection and recovery, optimal checkpointing intervals), and the HPC \emph{systems} (via efficient parallelization with various classical and modern programming models). 
To gauge the efficiency of additional modern programming paradigms (e.g., asynchronous tasking) against the classical ones for SPH, we also parallelize the SPH mini-app using classical paradigms, such as MPI, OpenMP, CUDA, and the more recent OpenMP target offloading, OpenACC, and HPX. 
This way, we cover all aspects that influence the performance and scalability of the SPH codes on modern and future HPC architectures.

\begin{figure}
    \centering
    \includegraphics[scale=0.36]{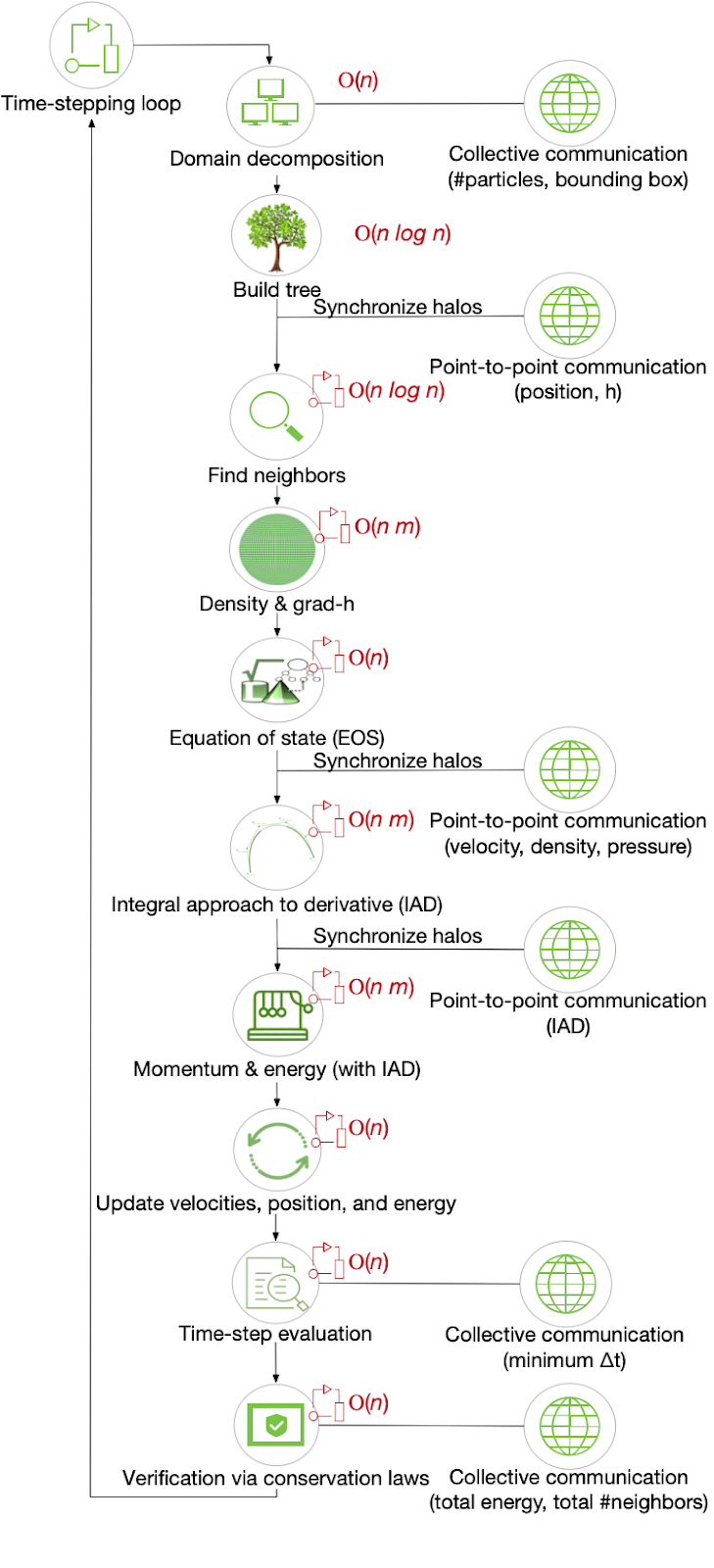}
    \caption{SPH general workflow. Computational steps are performed for every particle in the simulation domain over a number of time-steps. Point-to-point and collective communications update the particles information after certain computational steps. The complexity of each step is shown in red, where $n$ is the total number of particles and $m$ is the number of neighboring particles of each particle and depends on the smoothing length $h$.}
    \label{fig.workflow}
\end{figure}

In the first development stage, the focus has been on identifying and implementing the vanilla SPH solver (i.e. that with the original equations of \cite{monaghan1983}), the general workflow of which is shown in Figure~\ref{fig.workflow}. The solver has been designed from scratch as a distributed application. To achieve maximum efficiency and scalability on current Petascale and upcoming Exascale systems, one of the main challenges is to minimize the inter-process communication and synchronization. In particular, global synchronizations are avoided as much as possible as this would result in all processes having to communicate with all others, resulting in global idleness and loss of efficiency. Instead, we focused on developing a method that favors nearest neighbors communication between computing nodes, and only relies on collective communication for simple, yet necessary operations (e.g. computing the new size of the computational domain or the total energy).
Currently, the \sphexaminiapp includes a state-of-the-art implementation of the SPH equations (see Section.~\ref{subsec:sph}), which has been built progressively atop the optimal initial version.

\section{The \sphexaminiapp}

The \sphexaminiapp is described in the following, together with its latest developments, supported physics features, and parallel programming models. The code is open-source and is freely available on GitHub under the MIT license at: \url{https://github.com/xxx/yyy}.

\subsection{SPH}
\label{subsec:sph}
The SPH method has been implemented in the \sphexaminiapp following the formalism described in \cite{SPHYNX}. The main equations that express the calculation of the local density, and momentum and energy rates of change are:

\begin{align}
\rho_a =& \sum_b m_b W_{ab}(h_a)\,,\\
\left(\frac{dv_i}{dt}\right)_a =& -\sum_b m_b\left(\frac{P_a}{\Omega_a\rho_a^2}\mathcal A_{i,ab}(h_a)+\frac{P_b}{\Omega_b\rho_b^2}\mathcal A_{i,ab}(h_b)\right)+a_{i,a}^{AV}\,,\\
 \left(\frac{du}{dt}\right)_a =& \frac{P_a}{\Omega_a\rho_a}\sum_b\sum_i m_b(v_{i,a}-v_{i,b})\mathcal{A}_{i,ab}(h_a)+\nonumber\\
& \frac{1}{2}\sum_b\sum_i (v_{i,a}-v_{i,b})~a_{i,a}^{AV}\,,\\
a_{i,a}^{AV} =& \frac{1}{2}\sum_b m_b\Pi'_{ab}\left\{\frac{\mathcal A_{i,ab}(h_a)}{\rho_a}+ \frac{\mathcal A_{i,ab}(h_b)}{\rho_b}\right\}\,,
\end{align}

\noindent
where subindex $a$ is the particle index, $b$ runs for its neighbors indexes, and $i$ is the spatial dimension index. $\rho$, $m$, $P$, and $v_i$ are the density, mass, pressure, and velocity components of the particle, respectively. $W_{ab}$ is the SPH interpolation kernel, which depends on the local spatial resolution $h$, named smoothing length. $\Omega$ are the grad-h terms that take into account the changes in the local smoothing length. $\mathcal{A}_{i,ab}(h_a)$ are the terms for integral approach to derivatives, IAD, (see \cite{IAD, SPHYNX} for more details), and $a_{i}^{AV}$ are the artificial viscosity acceleration components, where:

\begin{equation}
\Pi'_{ab}= \begin{cases}
-\frac{\alpha}{2}~v_{ab}^{sig}~w_{ab} & \text{for {\bf x}$_{ab}\cdot${\bf v}$_{ab} < 0$,} \\
 0 & \text{otherwise,}
\end{cases}
\end{equation}

\noindent
is the artificial viscosity disipation term. $v_{ab}^{sig}=c_a+c_b-3w_{ab}$ is an estimate of the signal velocity between particles $a,b$, $c$ is the local speed of sound, and $w_{ab}= {\bf v}_{ab}\cdot {\bf x}_{ab}/\vert {\bf x}_{ab}\vert$. Finally, $\rho_{ab}^{-1}=2~(\rho_a+\rho_b)^{-1}$.

Updates to the position and velocity of the particles are done with a $2^{nd}$ order Press method, while energy is updated via a $2^{nd}$ order Adams-Bashforth scheme. 

The \sphexaminiapp currently implements the Sinc-family of interpolating kernels, the benefits of which are described in~\cite{SINC}:

\begin{equation}
W^s_n(v,h,n)=
\begin{cases}
\frac{B_n}{h^d}S_n(\frac{\pi}{2}v) &\text{for $0 \le v \le 2$,}\\
0 &\text{for $v > 2$,}
\end{cases}
\end{equation}
\label{eq:sinc}
where $S_n(.) = [sinc(.)]^n$, $n$ is a real exponent, $B_n$ a normalization constant, and $d$ the spatial dimension. The function $sinc$ is defined as: $sinc(\frac{\pi}{2}v)= sin(\frac{\pi}{2}v)/\frac{\pi}{2}v$, where $v=\vert{\bf x}_a-{\bf x}_b\vert/h_a$, and it is widely used in spectral theory.

Additional SPH kernels can be plugged in, upon implementation (in the above cases in only 6 LOC), proving how such a mini-app can be useful in allowing easy and quick modifications to experiment with alternative solutions. While not all SPH existing techniques and algorithms need to be implemented, a number of them, such as the SPH interpolation kernels, artificial viscosity treatments, or generalized volume elements, for example, can be later developed as separate interchangeable code modules. The \sphexaminiapp kernels have been optimized to enable excellent automatic vectorization via compiler optimizations.

Finally, as astrophysical and cosmological scenarios are within the scope of the \sphexaminiapp, we also implemented a multipolar expansion algorithm to evaluate self-gravity using the same tree structure that we use to find neighboring particles and based on which we perform the domain decomposition (see Section~\ref{subsec:domain-decomp} for more details on the latter).

These components are common to many production SPH codes and represent the basis over which we can optimize and extend the functionality of the \sphexaminiapp.

\subsection{Parallel Software Development}

The SPH workflow illustrated in Figure 1 is implemented in code as shown in the sequence diagram included in Figure 2. 
To ensure code flexibility, extensibility, and readability we follow solid principles of the code design, use continuous integration to avoid regression, and name functions and classes appropriately. Additionally we defined coding standard for the project which can be applied automatically by the use of the clang-format tool.

Regarding core-level optimization, particles are kept in memory in an order that matches the octree, so that particles that are close to each other in the 3D space, are also close in memory to minimize cache misses. Additionally, widely used values are precomputed and stored either in memory or in lookup tables.

The goal for developing \sphexaminiapp using parallel programming is to provide a reference implementation in MPI+X. 
The MPI standard is the \emph{de facto} communication library for distributed applications in HPC, due to the lack of an outperforming alternative for inter-node communication. 
OpenMP is the \emph{de facto} standard for parallel multithreaded programming on shared-memory architectures, widely used in academic, industry, and government labs. 
The hybrid MPI+OpenMP programming model represents a solid starting point for the parallel and distributed execution of the \sphexaminiapp. 
However, the vanilla MPI+OpenMP does not fully exploit the heterogeneous parallelism in the newest hardware architectures, 
Therefore, since version 4.5, the OpenMP standard~\cite{OpenMP} supports offloading of work to target accelerators.
Other languages directly targeting accelerators have been proposed and accepted by the community, such as OpenACC (a directive-based programming model targeting a CPU+accelerator system, similar to OpenMP), CUDA (an explicit programming model for GPU accelerators) and OpenCL.

The \sphexaminiapp currently implements HPX as an experimental development branch to explore the efficiency of task-based models and potential on (pre-)Exascale machines. The aims of exploring such task-based asynchronous programming model are to overlap computations and communications, both intra-node and inter-node, and to remove all synchronizations and barriers.

In terms of I/O from/to file system, the \sphexaminiapp has been designed from scratch to handle distributed data. The idea is that each computing node generates or loads a subset of the data. 
We currently support MPI I/O (in a branch) to perform parallel I/O operations at large scale by reading input and writing output binary data. We plan to move to HDF5 in the next months.

In terms of precision, round-off errors in single precision can add up to levels that might render the calculation useless, even using code units, mostly due to lack of accuracy in the evaluation of the gravitational forces using the tree. Nevertheless, this is based on the experience with the parent codes and we still have not studied this option in detail. Thus, we continue to use double precision data types while planning an exploration of a mixed-precision approach in future work.

\begin{figure*}
    \centering
    \includegraphics[scale=0.36]{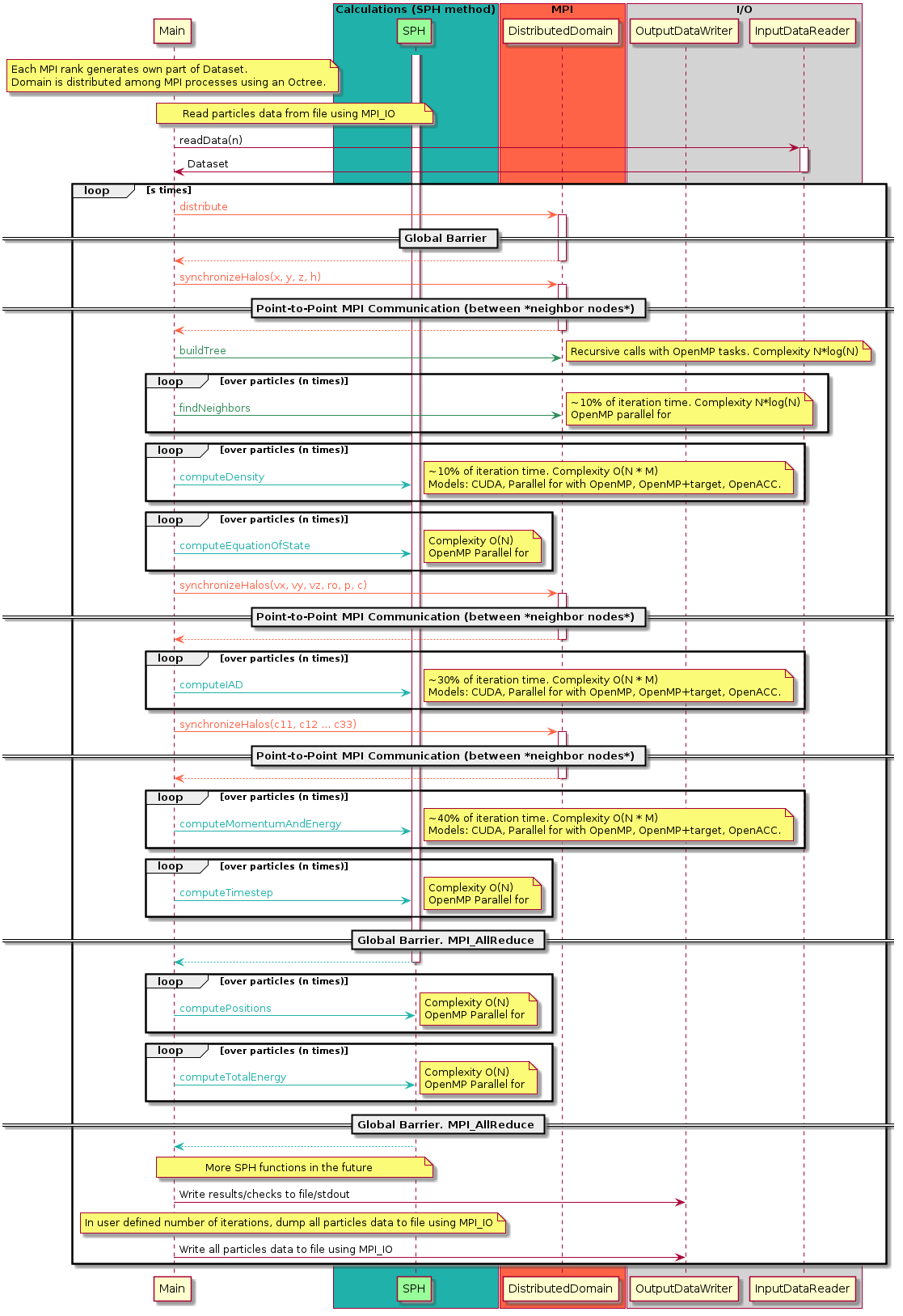}
    \caption{Sequence diagram of the \sphexaminiapp illustrating its parallel program flow, the use of the various parallel programming models, its complexity, and its synchronization points.}
    \label{fig.parallel.workflow}
\end{figure*}

The parallel program flow of the \sphexaminiapp for a typical test case (Rotating Square Patch~\cite{SqPatch}) is shown in Figure~\ref{fig.parallel.workflow}. 
The diagram describes the main execution loop, which is used to compute a time-step, and the sub-loops, where OpenMP and GPU offloading is used to accelerate the computation within single time-steps. 
Note that there are only three global synchronizations (\texttt{MPI\_AllReduce}) across distributed-memory nodes. The first synchronization is used to count the number of tree nodes when performing domain decomposition (described later in \S\ref{subsec:domain-decomp} and illustrated in Figure~\ref{fig.domain.decomposition}), the second synchronization is used to compute the minimum time-step ($\Delta t$) needed to advance the system in time, and the last synchronization is optional but useful for tracking total momentum and energy for verifying that they are conserved.

\subsection{Domain Decomposition}
\label{subsec:domain-decomp}

The spatial 3D domain of particles is decomposed into cells using an oct-tree data structure. The oct-tree is global — every node keeps an identical copy of it —  and it is created \emph{only once} at the beginning of the execution. The tree is then simply updated every iteration: branches are added or removed from the tree as needed. This global tree only contains general information such as the number of particles per cell, and does need to be refined down to single particles. Only the ‘top’ part of the tree is \emph{virtually shared} and maintained identical by every computing node. Compared to existing methods, this approach only requires two collective communications (\texttt{MPI\_AllReduce}): one to compute the number of particles in every cell of the tree and one to compute the new size of the spatial domain after particles have moved. Nevertheless, we aim at also circumventing these two collective communications, whose final objective is to rebuild the global tree to maintain particle load balance. Because particles can only move within the range of their local smoothing length, the upper levels of the global tree change at a much longer timescale than that of the particles dynamics. Therefore, global communications can be done scarcely while still avoiding particle load imbalance.

\begin{figure}[!htbp]
    \centering
    \vspace{-0.3cm}
    \includegraphics[scale=0.8]{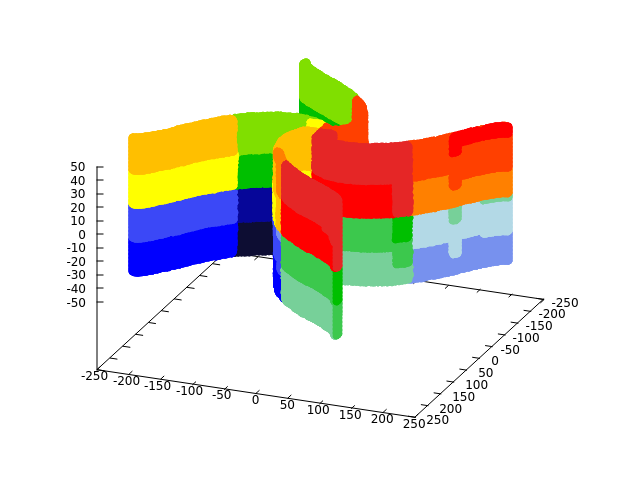}
    \caption{Domain decomposition for the Square Patch test case at time iteration 8,000 for 1M particles on 20 processes. Each color corresponds to a different MPI process.}
    \vspace{-0.25cm}
    \label{fig.domain.decomposition}
\end{figure}

Domain cells (tree branches) are assigned to MPI processes using the global tree to guarantee data locality, as shown in Figure~\ref{fig.domain.decomposition}.
The assignment process goes down to nodes that do not have particles fewer than the value globalBucketSize (a user-defined control parameter) which is typically equal to 128 particles. This way, certain processes can be assigned an additional tree node compared to others, but the difference in particles per process will be no more than the value assigned to globalBucketSize (as mentioned above). This causes imbalance equal to globalBucketSize/noOfParticlesPerMpiRank * 100\%, i.e., for 1 million particles per MPI rank and globalBucketSize of 128, the particle assignment imbalance will be equal to 0.01\%.

In practice, the imbalance is higher because MPI ranks will have different numbers of halo particles, i.e.  neighbors of an SPH particle located in the memory of another node need to be communicated and stored for the current iteration. Note that the amount of work does not increase (the number of neighbors remains the same, e.g. 300 per particle). However, more memory is required to store the additional particles. For example, the amount of halo particles for 300 neighbors varies by as much as 100\% for a few thousand particles per MPI rank to 1-5\% for a few million particles per MPI rank.

In addition, particle data is reordered in the local memory of every computing node to follow the depth-first-search ordering sequence of the oct-tree (similar to a Morton ordering sequence, or to using space filling curves, SFC, in general). This ensures that particles that are close together in the spatial domain are also close together in memory, resulting in increased cache efficiency.
In terms of memory consumption, 1,459B are needed per particle at ~1.35 GB per 1 Million particles and assuming 300 neighbors per particle. We use double precision floating point numbers for all physical properties and integers for storing particle’s neighbor indices. Currently we store the indices of the neighbors for each particle. 

\subsection{Communication}

\begin{figure*}[!htbp]
    \centering
    \includegraphics[width=1.0\textwidth]{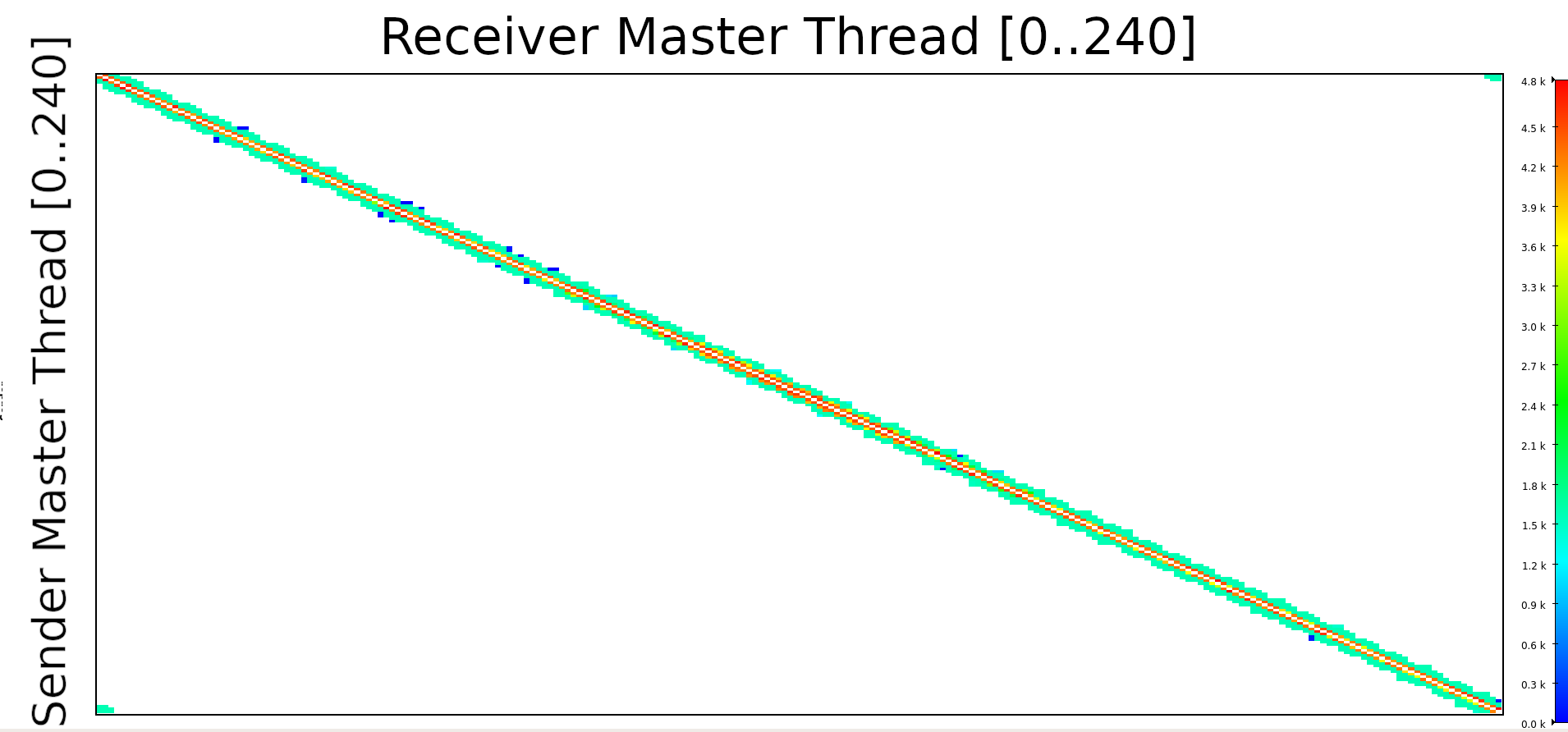}
    \caption{Communication between individual processes for a 240 MPI processes execution of the \sphexaminiapp with sender ID on the Y-axis and receiver ID on the X-axis.}
    \label{fig.communication}
\end{figure*}

Particles are exchanged between computing nodes in every simulation time-step when they move from one sub-domain to another. Halo particles are exchanged thrice per-time-steps. The \sphexaminiapp relies on asynchronous point-to-point communications (\texttt{MPI\_Isend}/\texttt{MPI\_Irecv}) between nodes to avoid global synchronizations. Figure~\ref{fig.communication} shows the communication matrix of the \sphexaminiapp for an execution using 240 MPI processes. This illustration shows that processes only communicate with their close neighbors, reducing latency and contention compared with naive collective-communication based implementations, while nodes remain synchronized with their neighbors in a loose fashion.

\subsection{Hybrid Computing}

The \sphexaminiapp offloads the most compute-intensive kernels (density, IAD and momentum equations) to GPUs. It can be compiled with or without support for multi-processing via MPI, multi-threading via OpenMP, and acceleration via OpenMP 4.5, OpenACC or CUDA. It supports the following combinations:
\begin{itemize}
\item MPI + OpenMP
\item MPI + OpenMP + OpenMP 4.5 target offloading
\item MPI + OpenMP + OpenACC offloading
\item MPI + OpenMP + CUDA 
\end{itemize}

\begin{figure}[htbp]
    \centering
    \includegraphics[scale=1.0]{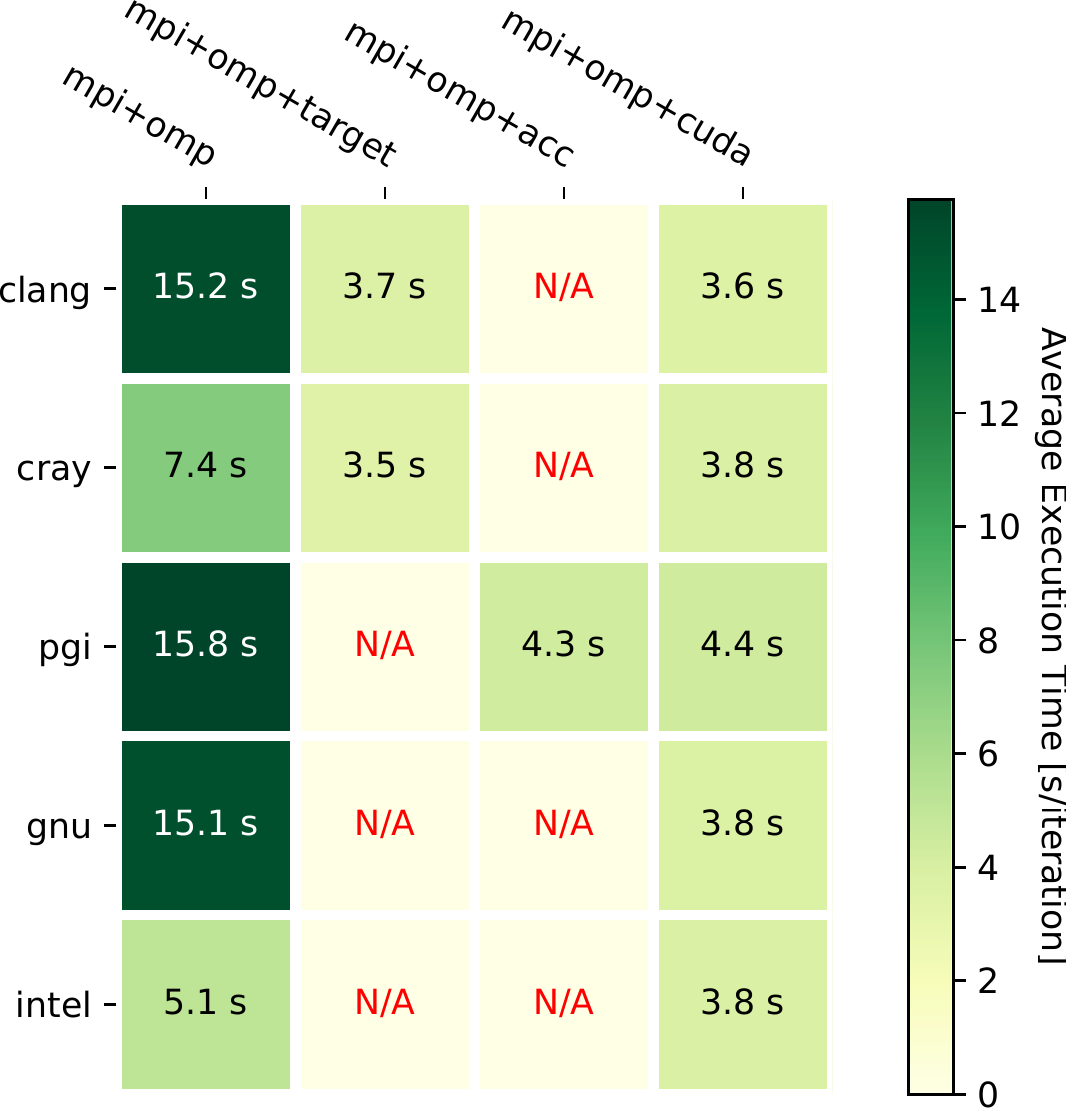}
    \caption{Average execution time per time-step for 5 compilers and for every parallel programming model available in the \sphexaminiapp.}
    \label{fig.hybrid}
    \vspace{-0.35cm}
\end{figure}

Figure~\ref{fig.hybrid} shows the average execution time for a single time-step using 8 million particles on 4 hybrid CPU+GPU nodes comprising 12 cores (Intel Xeon E5-2695) and \emph{a single} GPU card (NVIDIA Tesla P100), compiled using 5 different compilers and -O2 (or equivalent) optimization flag, for each available parallel model. Some models are not available (denoted N/A) on all compilers, either because they are explicitly not supported (e.g., the Intel compiler does not support OpenACC offloading) or because the compilers were not configured to support offloading at the time of writing.

Trigonometric functions are part of the SPH interpolation kernel (see Section~\ref{subsec:sph}), which is one the most commonly called function in every SPH iteration. As a result, math routines such as $cos()$, $sin()$ and $pow()$ are heavily used in the the mini-app.
Cray and Intel provide highly optimized math libraries with their compiler (the CSML and the MKL, respectively), which overrides the standard math function implementations and leads to inconsistencies in terms of runtime between the different compilers for the MPI+OMP model.
Note that this issue is not present with GPU offloading and CUDA, which both rely on Nvidia's CUDA implementation of these functions, hence the consistent runtimes for the other models.

For this reason, we have implemented a lookup table which contains 20,000 precomputed kernel values and we perform a linear interpolation with with the relative distance between neighboring particles at runtime. This avoids calls to complex and/or costly mathematical functions. In addition, most calls to $pow(x,n)$, rely on an integer $0 \leq n \leq 9$ and so we can replace each call by the corresponding inline $x*x*x*x$... which significantly outperforms the standard implementation for the GCC, Clang and PGI compilers.
Overall, these optimizations lead to similar and consistent execution times for all compilers, comparable to using the Intel or Cray compiler (not shown Figure~\ref{fig.hybrid}).

\subsection{Verification and Validation}
\label{subsec:vnv}
The complexity of the scenarios simulated in CFD and Astrophysics usually prevent the possibility of performing simulations with continuously increased resolution and different codes, so that a convergence to zero differences on the results can be found. 
Often, it is neither possible nor reasonable to obtain sufficient computational resources to perform simulations that are ``converged'' throughout the computational domain in a mathematical sense. 
It is much more important to limit the deviations in under-resolved regimes by enforcing fundamental conservation laws. As a consequence, overall physics properties of the simulated scenarios remain robust, even if slightly different results are obtained when using different codes to solve the same set of equations. 
Therefore, comparing results of different hydrodynamical codes to the same initial conditions has been proved to be highly beneficial to gain understanding in complex scenarios, in the behavior of the codes, and to discover their strengths and weaknesses. These comparisons are common in CFD and Astrophysics~\cite{liebendoerfer2005, agertz2007, tasker2008, cabezon2018}.

The common test case chosen to validate the results obtained through the mini-app against the ones of the parent codes is the Rotating Square Patch. This test was first proposed by~\cite{SqPatch} as a demanding scenario for SPH simulations. The presence of negative pressure stimulates the emergence of tensile instabilities of numeric origin that induce unrealistic clumping of particles. This leads to an unphysical evolution of the fluid and ultimately to the interruption of the simulation. Nevertheless, these can be suppressed either using a tensile stability control or increasing the order of the scheme~\cite{SPHFLOW}. As a consequence, this is a commonly used test in CFD to verify hydrodynamical codes, and it is employed in this work as a common test for the three parent codes.

The setup here is similar to that of \cite{SqPatch}, but in 3D. The original test was devised in 2D, but the SPH codes used in this work normally
operate in 3D. To use a test that better represents the regular operability of the target codes, the square patch was set to $[100 \times 100]$ particles in 2D and this layer was copied 100 times in the direction of the $z$-axis. This results in a cube of $10^6$ particles that, when applying periodic boundary conditions in the $z$ direction, is similar to solving the original 2D test 100 times\footnote{Note that in the 3D case it is expected that instabilities arise in the Z direction. Nevertheless, in all our experiments, the velocity for all particles in the Z direction was, most of the time, 6 orders of magnitude lower than in the X or Y directions. Only for few particles close to the origin of coordinates (where all velocities are small) this ratio increases to $10^{-2}$. That is the point at which we stopped the SPH simulation. Hence, we consider the 3D test stable for the duration of the simulation and the results comparable to those found in the literature for the 2D case.}, while conserving the 3D formulation of the codes. The initial conditions are the same for all layers, hence they depend only on the $x$ and $y$ coordinates. The initial velocity field is given such that the square rotates rigidly:
\begin{equation}
v_x(x,y)=\omega y;\qquad v_y(x,y)=-\omega x,
\label{eq:square_v0}
\end{equation}
\noindent
where $v_x$ and $v_y$ are the $x$ and $y$ coordinates of the velocity, and $\omega = 5$~rad/s is the angular velocity. 
The initial pressure profile consistent with that velocity distribution can be calculated from an incompressible Poisson equation and expressed as a rapidly converging series:
%
\begin{multline}{\notag}
P_0=\rho\sum_{m=0}^\infty\sum_{n=0}^\infty\frac{-32\omega^2}{mn\pi^2\left[\left(\frac{m\pi}{L}\right)^2 + \left(\frac{n\pi}{L}\right)^2 \right]} \times  \\
\sin\left(\frac{m\pi x}{L}\right)\sin\left(\frac{n\pi x}{L}\right),
\end{multline}
\label{eq:square_p0}
%
\noindent
where $\rho$ is the density and $L$ is the side length of the square.


We also verified the results of the \sphexaminiapp against the outcome of simulating the same rotating square patch test, with the same initial conditions, in all three parent codes. To that extent we compared the total angular momentum $\vec{L}_{tot}$ at $t=0.5$~s among all codes. The mini-app yields $\vec{L}_{tot}=8.33\times 10^9$~g$\cdot$cm$^2$, which differs from the parent codes by $\sim 0.2\%$. This is well within the differences among the parent codes themselves. This small discrepancy comes from the different implementation of relevant sections of the SPH codes. Namely, the way of calculating gradients, volume elements, evaluate the new time-step, and integrate the movement equations. Nevertheless, despite these capital differences in implementation among the codes, the results converge well enough to ensure an adequate evolution of the system, pointing at the fact that the fundamentals of the SPH technique are correctly implemented in the \sphexaminiapp.

\section{Experiments}

In this section, we report the results of a weak-scaling and a strong-scaling experiment conducted on a top production supercomputer. The immediate goal is to assess the performance of the \sphexaminiapp on a Petascale system comprising both CPUs and GPUs. The long-term goal is to run on Exascale systems. 
The weak-scaling efficiency is obtained as $T_{seq} /T_{par} \times 100\%$. For strong-scaling, we plotted the average wall-clock time per iteration on a logarithmic scale. 
The average time per iteration is obtained from 10 iterations due to the limited number of node hours available for this project. However, a single run up to 8000 iterations used for validation has shown that the time per execution remains almost constant over 10 iterations with very small variations.

We show that the proposed \sphexaminiapp shows promising results, executing simulations with 65 billion particles on 2,048 hybrid CPU+GPU nodes at 67\% weak-scaling efficiency. Moreover, the achieved weak-scaling efficiency decreased very little when scaling up from 512 to 2,048 nodes while keeping 32 million particles/node. 
For strong-scaling, the average time per iteration decreases almost linearly up to 1024 nodes for a fixed problem size of $644^{3} = 267M$ particles.
Memory consumption equals to 1,459B per particle, 1.35GB per 1 million particles.

In addition, an independent performance audit of the \sphexaminiapp has recently been performed by RWTH Aachen as part of the POP2 CoE service for European Scientific Applications.
The report analyzed the efficiency and scalability of the \sphexaminiapp through a set of strong-scaling experiments with up to 960 MPI ranks (40 nodes) and showed that both are very high.

\subsection{System Overview}

The experiments were performed on the hybrid partition of the \systemW\footnote{\url{https://www.cscs.ch/computers/piz-daint/}} 
supercomputer using PrgEnv-Intel, Cray MPICH 7.7.2 and OpenMP 4.0 (version/201611).

We used the hybrid partition of more than 5,000 Cray XC50 nodes. These hybrid nodes are equipped with an Intel E5-2690 v3 CPU (codename Haswell, each with 12 CPU cores) and a PCIe version of the NVIDIA Tesla P100 GPU (Pascal architecture, 3584 CUDA cores) with 16 GB second generation high bandwidth memory (HBM2). The nodes of both partitions are interconnected in one fabric based on Aries technology in a Dragonfly topology\footnote{\url{http://www.cray.com/sites/default/files/resources/CrayXCNetwork.pdf}}.

\subsection{Weak-Scaling Experiments}

\begin{figure}[htbp]
    \centering
    \vspace{-0.15cm}
    \includegraphics[scale=0.6]{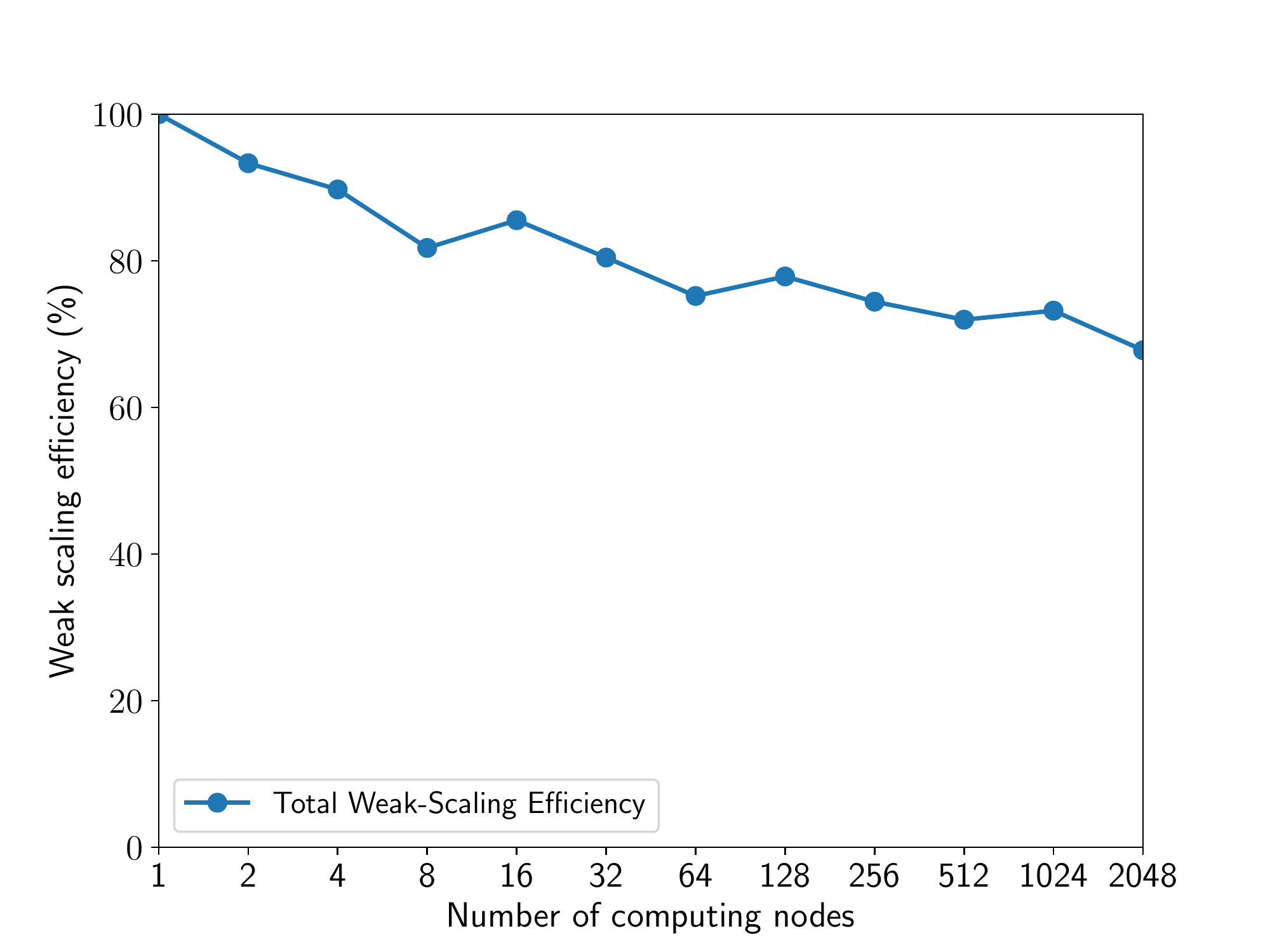}
    \caption{Weak scaling of the \sphexaminiapp with 32 million particles per node}
    \label{fig.weak-scaling}
    \vspace{-0.05cm}
\end{figure}

\begin{figure}[htbp]
    \vspace{-0.15cm}
    \centering
    \includegraphics[scale=0.6]{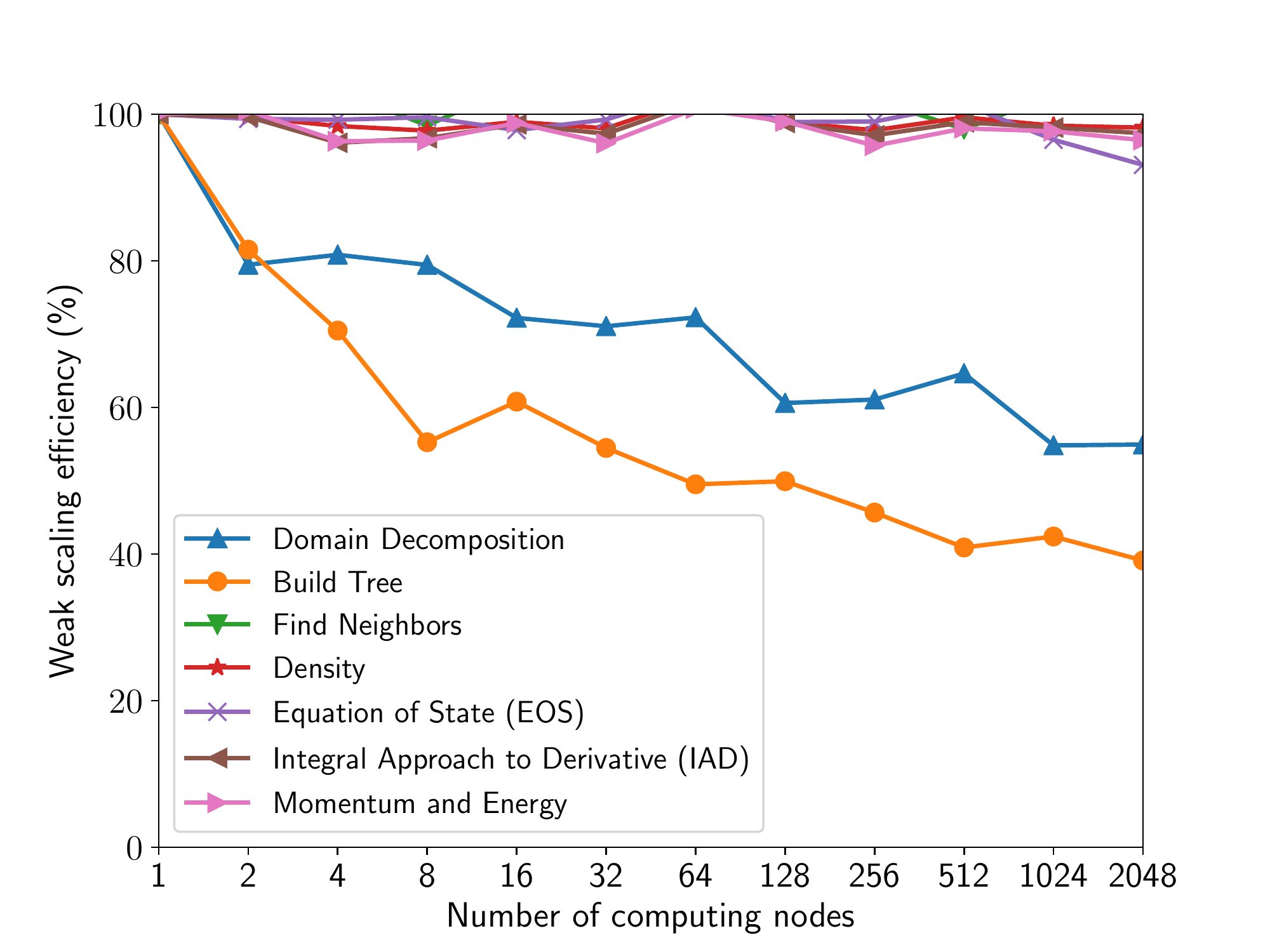}
    \caption{Weak scaling per function of the \sphexaminiapp with 32 million particles per node}
    \label{fig.weak-scaling-steps}
    \vspace{-0.15cm}
\end{figure}

Figure~\ref{fig.weak-scaling} and Figure~\ref{fig.weak-scaling-steps} show the results of a weak-scaling experiment conducted on the hybrid partition of the \systemW supercomputer. Both figures show the efficiency of the execution with increasing number of nodes relative to using a single computing node. The run performs the same 3D version of the CFD rotating square patch test described in $\S$\ref{subsec:vnv}, with 32 million particles per computing node for a total of 65 billion particles on 2,048 nodes. The \sphexaminiapp shows very good parallel efficiency, namely 67\% when running the largest test case. Note that the mini-app only experienced a 3\% decrease in parallel efficiency when moving from 512 nodes to 2,048 nodes, i.e., increasing the number of MPI ranks and the problem size, by a factor of 4$\times$.
\label{subsec:scaling}

Figure~\ref{fig.weak-scaling-steps} shows the breakdown of efficiency by function. We observe that the decrease in efficiency is mostly due to the Domain Decomposition and Build Tree steps. Domain Decomposition is the most complex part of the code, responsible for distributing the particle data and performing load balancing. While the number of particles per process remains constant, the global tree -- the top part of the tree that is kept identical on all processes -- becomes larger, and this results in increased depth and an additional overhead as the number of nodes (and particles) increases globally. However, the depth of the tree increases with $log(n)$, where $n$ is the total number of particles. This is reflected in Figure~\ref{fig.weak-scaling-steps}, orange and blue lines, where the efficiency decreases rapidly when increasing the node count from 1 to 64, but then remains stable and decreases very little (note the logarithmic scale on the X-axis). All other computing steps scale almost perfectly.

These results are very encouraging and indicate that good scalability can be expected at much higher node counts. The \sphexaminiapp has not yet reached the tipping point where the code will not benefit from increased computing resources. If we assume the current efficiency trend, we can expect the code to run in excess of a trillion particles on a system consisting of 31,250 computing nodes with 32 million particles per nodes, which will be on par with future Exascale systems node counts. 

\subsection{Strong Scaling Experiments}
\label{subsec:strongscaling}

\begin{figure}[htbp]
   \vspace{-0.15cm}
    \centering
    \includegraphics[scale=0.6]{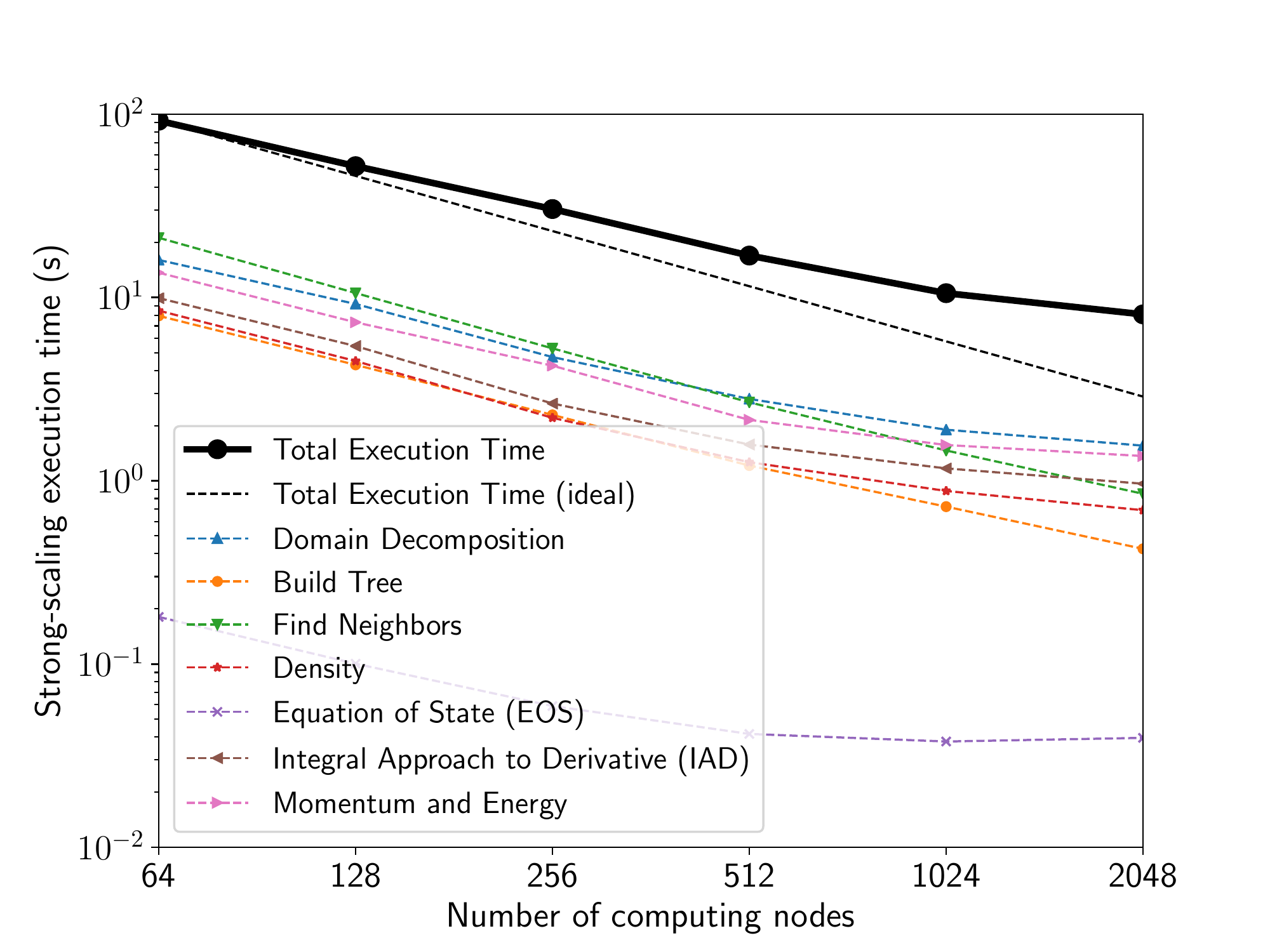}
    \caption{Strong scaling of the \sphexaminiapp with 267 million particles}
    \label{fig.strong-scaling}
    \vspace{-0.15cm}
\end{figure}

Figure~\ref{fig.strong-scaling} shows the results of a strong-scaling experiment conducted with the \sphexaminiapp on 267 million particles on the hybrid partition of the \systemW supercomputer using up to 2,048 nodes. 
The results were obtained by running the mini-app \ac{using MPI+OpenMP+CUDA. Additional dotted lines show the execution time per function of the mini-app.}



\ac{
While the \sphexaminiapp has a sub-optimal speedup, it is clear that all functions benefit from the additional computational resources, and that the execution time continues to decrease even when using 2,048 nodes.
In this strong-scaling scenario the total number of particles is fixed and nodes receive fewer particles as we increase the number of computing nodes, e.g., there are 4.1 million particles per node with 64 computing nodes, but only 130,000 per node (i.e. just above 10,000 per core) when using 2,048 computing nodes. 
However, particles maintain the same number of neighbors. This means the overlap between nodes increases and more halo particles are needed.}

\begin{figure}[htbp]
    \vspace{-0.15cm}
    \centering
    \includegraphics[scale=0.6]{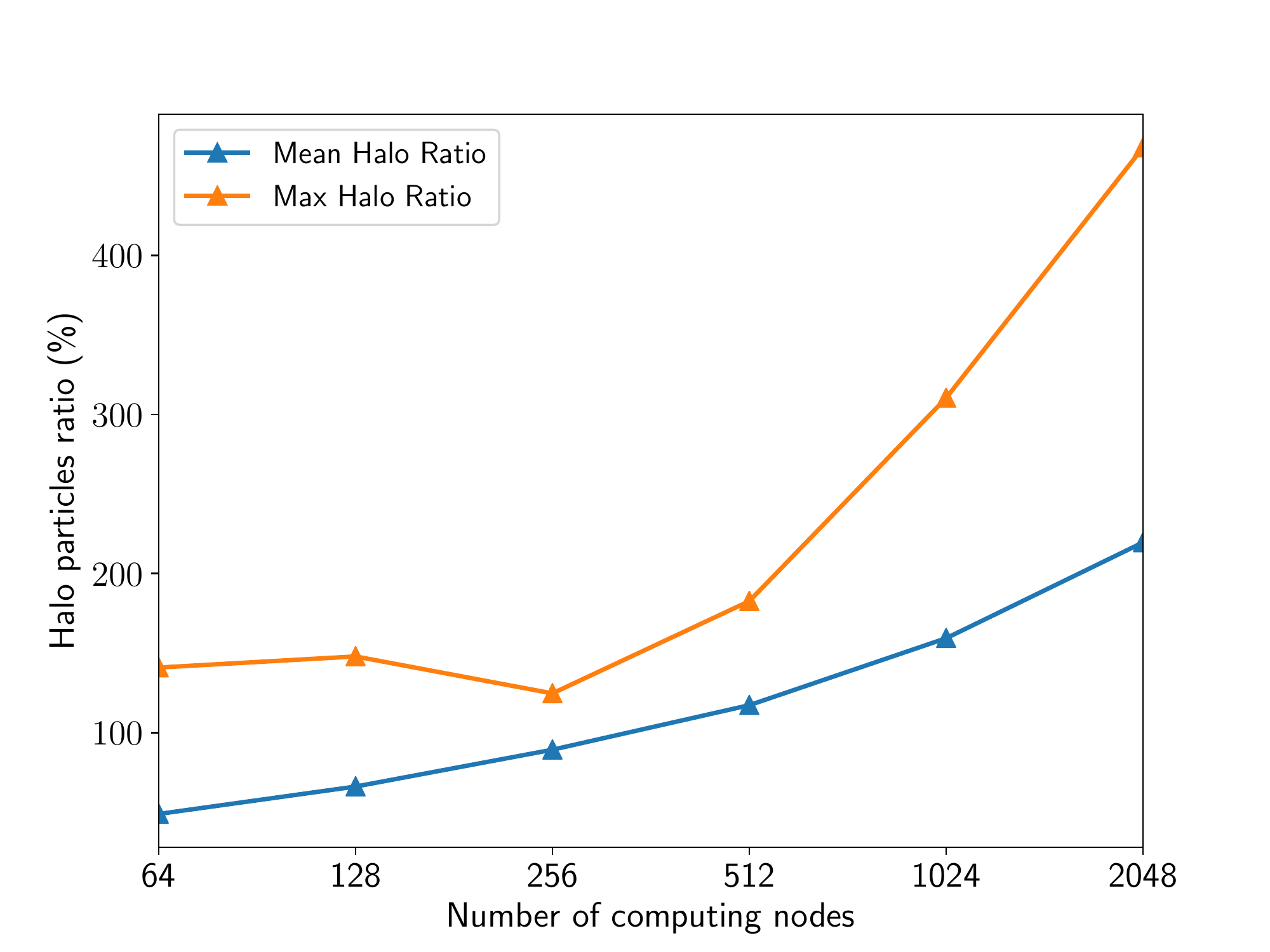}
    \caption{\textit{Mean} and \textit{max} ratio of halo particles per computing node w.r.t. initially assigned particles per node, for the test case with 267 million particles.}
    \label{fig.haloRatio}
    \vspace{-0.15cm}
\end{figure}

\ac{Figure~\ref{fig.haloRatio} shows the \textit{mean} and \textit{max} ratio of halo particles per node, with respect to the number of particles initially assigned to a node, as described in Section~\ref{subsec:domain-decomp}.
While only 30\% of extra halos particles are needed on average using 64 nodes, more than 200\% are needed when using 2,048 nodes.}

\ac{The impact of having more halo particles per node is two-fold: (1) additional memory is required and (2) particles are more likely to be distant in memory space, which increases the number of cache-misses. Both aspects adversely impact performance.
In a realistic setup, i.e., from 64 to 128 compute nodes in Figure~\ref{fig.haloRatio}, we aim at a few million particles per nodes and a few hundred thousand particles per core / thread. Such configurations also minimize the number of halo particles per node.}

\subsection{Current Limitations}

The weak-scaling experiment used $100\%$ of the available memory on every computing node and on every GPU. The execution time for a single time-step remained, on average, well under a minute. Hence, a high priority optimization is to reduce the memory footprint of the \sphexaminiapp, which will come at the expense of longer execution times but will allow us to address and execute larger problem sizes in the future.

Our current project allocation on \systemW did not allow us to run larger experiments at the time of writing. We are currently working on obtaining larger allocations to launch simulations using the full system.


\section{Next Steps and Future Work}

The mini-app provides a modular tool that can be used by the community to plug-in additional physical processes or numerical methods. As there is a wide variety of very demanding physical scenarios that will be targeted by the \sphexaminiapp (e.g. supernova explosions, galaxy formation), having a highly scalable and modular code will encourage others in the community to contribute with their own physics/numerics modules.

Our short-term objectives include reducing the memory footprint of the \sphexaminiapp, which will help increasing the particle count per node beyond 32 million particles, adding additional features such as the possibility to select the desired generalized volume elements, and simulating increasingly complex scenarios. Regarding fault tolerance, we plan to include our novel detection method for silent data corruption~\cite{CavelanCC19}, which is specifically designed for SPH applications, add automatic validation (through conserved quantities), and implement checkpointing at optimal intervals. Each new feature in the \sphexaminiapp will be iteratively tested, validated, and optimized for efficiency from the start.

As mid-term targets we aim to overlap computations with inter-node communications. Therefore, we will prepare the \sphexaminiapp to be Asynchronous Multi-Tasking (AMT) ready. For this, we will enable the opportunity to compare different tasking frameworks such as OpenMP and HPX. Moreover, this will also open the opportunity to delegate independent tasks to accelerators and CPUs at the same time, which, in terms of the increasing hardware heterogeneity foreseen in the pre-exascale and exascale systems, is crucial to achieve maximum load balance and performance. Additionally, we will employ multilevel (batch, process, and thread) scheduling for dynamic load balancing in the mini-app as a configurable option.
This will allow us to systematically explore the interplay between load balancing at these levels to achieve the best possible load balancing during execution.

In the end, the \sphexa follows a long-term vision, having set out to have major impact in the scientific communities it gathers (and beyond in the longer run). The aim of \sphexa is to reach the capabilities of present HPC systems and to push those of the future HPC infrastructures for simulating the most complex phenomena at the highest resolution and longest physical times, such as exploring the explosion mechanisms of Type Ia and Core Collapse Supernovae, including nuclear reaction treatments via efficient nuclear networks and neutrino interactions with detailed transport, respectively. Another target for the \sphexaminiapp is modeling the small-scale fluid-dynamics processes involved in the assembly of the planetary building blocks while capturing simultaneously the large scale dynamics of the proto-planetary disks, and being able to simulate an entire population of galaxies from high to low redshift in a cosmological volume with enough resolution to model directly individual star forming regions.
Hence, a flexible and modular code that scales efficiently and robustly on a large number of nodes nodes, accessing a mix of CPUs, GPUs, FPGAs, etc, is our long-term objective.

\section{Conclusion}

\sphexa is an interdisciplinary project involving Computer Scientists,  Astrophysicists, and Cosmologists, brings together state-of-the-art methods in both fields under a broad scope, and addresses important challenges at all levels of existing and emerging infrastructures by exploring novel programming paradigms and their combinations.

In this work, we described the current status of a novel and scalable \sphexaminiapp for simulating the SPH method on large hybrid HPC systems efficiently utilizing both multi-core CPUs and GPUs.
The \sphexaminiapp is open source and has no external dependencies. With fewer than 3,000 modern C++ LOC (with no compromise for performance), it is easy to run, understand, modify, and extend. The code is simple by design, making it easy to test and implement new SPH kernels or other performance optimizations.
We performed an initial exploration of the efficiency of different combinations of hybrid CPU and GPU programing models, with different compilers, verified and validated the \sphexaminiapp via computationally-demanding simulations, and compared the results with those obtained by the parent codes.
We also conducted a weak-scaling experiment that shows excellent scaling, with 67\% efficiency on 2,048 hybrid nodes of the \systemW supercomputer with a loss of just 3\% in efficiency when going from 512 to 2,048 nodes.

At this initial stage of the project we are already in a good position to explore the limits of what can be done on current top supercomputers. This opens the doors to great number of potential applications, not only of the \sphexaminiapp, but also of the learned lessons. 
The \sphexaminiapp has driven substantial improvements to its three parent codes (\sphynx, \changa, and \sphflow) and has set a multi-directional knowledge transfer between the Computer Science, Astrophysics, and Computational Fluid Dynamics communities.


\section*{Acknowledgements}

This work is supported in part by the Swiss Platform for Advanced Scientific Computing (PASC) project SPH-EXA~\cite{SPHEXA} (2017- 2021).
The authors acknowledge the support of the Swiss National Supercomputing Centre (CSCS) via allocation project c16, where the calculations have been performed.
Several performance results were provided by the Performance Optimisation and Productivity (POP) centre of excellence in HPC.

\bibliographystyle{ACM-Reference-Format}


\end{document}